\documentclass[reprint,aps,pra,superscriptaddress,showpacs,
amsmath,amssymb]{revtex4-1}
\usepackage{graphics}
\usepackage{mathtools}
\usepackage{graphicx}
\usepackage[version=3]{mhchem} 
\usepackage[T1]{fontenc}       
\usepackage{color}
\usepackage{MnSymbol}
\usepackage{tikz}
\usepackage{amsfonts}
\usepackage{amsmath}
\allowdisplaybreaks

\newcommand{\pd}[2]{\frac{\partial #1}{\partial #2}} 
\newcommand{\bs}{\boldsymbol}
\newcommand{\bt}{\textbf}
\newcommand{\im}{\text{i}}

\newcommand{\be}[0]{\begin{equation}}
\newcommand{\ee}[0]{\end{equation}}
\newcommand{\bea}[0]{\setlength\arraycolsep{2pt}\begin{eqnarray}}
\newcommand{\eea}[0]{\end{eqnarray}}
\newcommand{\bse}{\begin{subequations}}
\newcommand{\ese}{\end{subequations}}
\newcommand{\bdm}[0]{\begin{displaymath}}
\newcommand{\edm}[0]{\end{displaymath}}

\newcommand{\bc}[0]{\begin{center}}
\newcommand{\ec}[0]{\end{center}}

\newcommand{\ud}[0]{\mathrm{d}}

\begin{document}

\title{Lorenz-Mie scattering of focused light via complex focus fields:
 an analytic treatment}
%
%

\author{R. Guti\'{e}rrez-Cuevas}
\email{rgutier2@ur.rochester.edu}
\affiliation{The Institute of Optics, University of Rochester, Rochester, NY 14627, USA}
\affiliation{Center for Coherence and Quantum Optics, University of Rochester, Rochester, NY 14627, USA}
\author{Nicole J. Moore}
\affiliation{Department of Physics, Gonzaga University, Spokane, WA 99258, USA}
\author{M. A. Alonso}
\affiliation{The Institute of Optics, University of Rochester, Rochester, NY 14627, USA}
\affiliation{Center for Coherence and Quantum Optics, University of Rochester, Rochester, NY 14627, USA}
\affiliation{Aix Marseille Universit\'e, Centrale Marseille, Institut Fresnel, UMR 7249, 13397 Marseille Cedex 20, France}




\begin{abstract}
The Lorenz-Mie scattering of a wide class of focused  electromagnetic fields off spherical particles is studied. The focused fields in question are 
constructed through complex focal displacements, leading to closed-form expressions that
can exhibit several interesting physical properties, such as orbital and/or spin angular momentum, spatially-varying polarization, and a 
controllable degree of focusing. These fields constitute complete bases that can be considered as nonparaxial extensions of the 
standard Laguerre-Gauss beams and the recently proposed polynomials-of-Gaussians beams. 
Their analytic form turns out to
lead also to closed-form expressions for their multipolar expansion. Such expansion can be used to compute the field scattered by a 
spherical particle and the resulting forces and torques exerted on it, for any relative position between the field's focus and the particle. 
\end{abstract}

%
%
%


\maketitle
\section{Introduction}

Optical trapping and manipulation constitute important techniques in research on systems that range from the biological to the quantum-
mechanical \cite{neuman2004optical, grier2003revolution,dholakia2011shaping}, with the size of the 
trapped object(s) varying by around four orders of magnitude. 
Single-beam optical traps (often known as optical tweezers) are made possible by the existence of the so-called gradient force. When 
the irradiance gradient is sufficiently large,  this force can counteract the radiation pressure exerted on the object by the field, as first 
determined by Ashkin and collaborators \cite{ashkin1986observation,ashkin2000history}.  Early analysis 
\cite{ashkin2000history,novotny2006principles} of single-beam optical traps focused on scatterers whose size is much smaller than 
the optical wavelength, $\lambda$, allowing the use of Raleigh's approximation in which the expressions for the induced forces and 
torques take simple forms.  Later work considered the opposite limit, where the size of the scatterer is much larger than $\lambda$ thus 
allowing an accurate treatment in terms of geometrical optics \cite{ashkin1992forces,volke-sepulveda2004three}.  

In order to achieve a sufficiently large gradient force to produce a stable trap, the incident light must be strongly focused, requiring a 
high numerical aperture microscope objective.  Analysis of the forces and torques acting on an object whose size is between the Rayleigh 
and geometrical optics limits requires a generalization of Lorenz-Mie scattering, which, in its original form, treated the scattering of a plane 
wave by a spherical particle. 
A substantial body of work has been devoted to generalizing Lorenz-Mie scattering to treat arbitrarily-shaped fields and scatterers (c.f., \cite{barton1988internal,barton1989theoretical,
gouesbet1982scattering,mishchenko2004t,mishchenko1999light,
nieminen2011t,gouesbet2017generalized,
gouesbet2015electromagnetic}).

Scattering of arbitrary fields is described using the generalized Lorenz-Mie theory (GLMT) \cite{gouesbet2015electromagnetic,gouesbet2017generalized}, 
which relies on the decomposition of the incident field in terms of vector multipoles \cite{sheppard1997efficient,jackson1998classical,borghi2005highly,
nieminen2003multipole} and the fulfillment of boundary conditions through the use of the appropriate T-matrix \cite{mishchenko2004t,mishchenko1999light,nieminen2011t,
gouesbet2017generalized}. Past treatments of the scattering of focused electromagnetic fields \cite{nieminen2003multipole,kiselev2014mie,kiselev2016optical,
 nes2007rigorous,yu2014radiation,yu2015radiation,
simpson2008rotation,simpson2009optical,simpson2010orbital} provide a variety of models for optical tweezers.  These models differ from each other primarily in how the focused incident field is described: through field matching \cite{nieminen2003multipole,kiselev2014mie,kiselev2016optical}, through use of the Richards-Wolf diffraction theory (c.f., \cite{richards1959electromagnetic,novotny2006principles,
dutra2014absolute}) to focus a paraxial beam by an optical element \cite{nes2007rigorous,yu2014radiation,yu2015radiation}, or via an ad-hoc extension of paraxial beams to the nonparaxial electromagnetic regime \cite{simpson2008rotation,simpson2009optical,simpson2010orbital,
barnett1994orbital}.
In all these treatments, the expressions for the coefficients of the vector multipoles in the beam decomposition are typically not analytic; i.e., computation of these expressions requires numerical integration.  

The present treatment relies on models for the focused incident field that allow for analytic expressions for the coefficients in the decomposition.  These models are based on what is referred to here as complex focus (CF) fields \cite{berry1994evanescent,sheppard1998beam,
sheppard1999electromagnetic,tagirdzhanov2015complexified} 
and their generalizations to complete sets of bases \cite{moore2009closed,gutierrez-cuevas2017scalar}. Please note that complex focus fields are similar to complex source-point (CSP) fields \cite{kravtsov1971complex,deschamps1971gaussian,
orlov2010complex,orlov2014vectorial,tagirdzhanov2015complexified} except that they are free of the branch ring at the focal plane that 
makes CSP fields singular multivalued solutions; CF fields are analytic everywhere.  CF fields are known to be a rigorous nonparaxial 
extension of Gaussian beams \cite{berry1994evanescent,sheppard1998beam,
sheppard1999electromagnetic} (required since Gaussian beams are a solution to the paraxial wave equation that is not valid in the high-
numerical aperture case).  Previous work on Mie scattering by CF fields has been limited to low order modes or to the scalar regime 
\cite{moore2008closed,lombardo2012orthonormal,
orlov2012analytical,
moore2016mie}.

In the present work, we describe the Lorenz-Mie scattering of the elements of the bases given in \cite{moore2009closed,
gutierrez-cuevas2017scalar}. The fields derived in \cite{moore2009closed} can be considered as rigorous nonparaxial 
electromagnetic analogues to the Laguerre-Gauss (LG) beams, given in terms of a closed-form expression, without the need for 
undetermined functions and numerical integrals \cite{barnett1994orbital,nes2007rigorous}, whereas those proposed 
in \cite{gutierrez-cuevas2017scalar} are the nonparaxial electromagnetic extension of the polynomials-of-Gaussians (PG) 
bases \cite{gutierrez-cuevas2017polynomials,gutierrez-cuevas2017complete}, which present some useful confinement 
properties while providing a radial and angular structure similar to that of LG beams. These bases allow the study of fields with orbital 
angular momentum (OAM) \cite{dholakia2011shaping,
yao2011orbital,andrews2012angular} and its coupling with polarization or spin angular momentum \cite{bliokh2010angular}. 
We derive analytic expressions for the coefficients in their multipolar expansion, valid for any relative position between the scatterer and 
the incident field. Therefore, the use of a translation equation with limited accuracy  
\cite{cruzan1962translational,mishchenko2002scattering} can be avoided.

\section{Vector multipoles}

We restrict our attention to monochromatic electromagnetic free fields in homogeneous transparent media, with assumed time 
dependence given by $\exp(-\im \omega t)$. These fields satisfy the vector Helmholtz equation,
\begin{align}
\label{eq:helm}
\nabla^2\textbf{E}(\textbf{r}) +k^2 \textbf{E}(\textbf{r})=0,
\end{align}
with the added divergence condition
\begin{align}
\label{eq:div}
\nabla\cdot \textbf{E}(\textbf{r}) =0,
\end{align}
where $k=2\pi n_0/\lambda$ is the wavenumber and $n_0$ is the index of refraction of the background medium, which is assumed to be real.
Any such field can be expressed as a continuous superposition of plane waves, that is, 
\begin{align}
\textbf{E}(\textbf{r})=\int_{4\pi}\textbf{A}(\textbf{u})e^{\im k{\bf u} \cdot {\bf r}} \ud \Omega,
\end{align}
where $\bt u$ is a real unit vector indicating the plane wave direction and  $\textbf{A}(\textbf{u})$ is the plane-wave amplitude (PWA), 
often referred to as the angular spectrum. Note that the divergence condition imposes transversality of the PWA, 
$\bt u \cdot \bt A ({\bf u})=0$. The unit vector $\bt u$  is the variable of integration through the angles $\theta$ and $\phi$ in PWA (or direction) space.

In spherical coordinates, Eqs.~(\ref{eq:helm}) and (\ref{eq:div}) yield the standard source-free solutions $\bs \Lambda_{l,m}^{(\text{I})}$ 
and  $\bs \Lambda_{l,m}^{(\text{II})}$ known as the vector multipoles (also referred to as the vector spherical wavefunctions and denoted 
as $\bt M_{l,m}$ and $\bt N_{l,m}$ \cite{gouesbet2015electromagnetic,gouesbet2017generalized}). They are given by
\bse
\label{eq:vectmult}
\begin{align}
\bs \Lambda^{(\text{I})}_{l,m}(\textbf{r})=& \frac{1}{k\sqrt{l(l+1)}}\nabla\times\nabla\times [ \textbf{r} \Lambda_{l,m}(\textbf{r})] ,\\
\bs \Lambda^{(\text{II})}_{l,m}(\textbf{r})=& \frac{\im}{\sqrt{l(l+1)}}\nabla\times [ \textbf{r} \Lambda_{l,m}(\textbf{r})],
\end{align}
\ese
where 
\be
\label{eq:scmult}
\Lambda_{l,m}(\textbf{r})=4\pi \im^l j_l(kr)Y_{l,m}(\theta_r, \phi_r)
\ee
are the scalar mutipoles, with 
\begin{align}
Y_{l,m}(\theta,\phi)=\sigma_m^m\sqrt{\frac{(2l+1)(l-m)!}{4 \pi (l+m)!}}P^{(m)}_l(\cos \theta)e^{\im m\phi}
\end{align}
being the spherical harmonics [$\sigma_m=\text{sgn}(m+1/2)$] and $j_l$  the spherical Bessel functions. Here, $\theta_r$, $\phi_r$ and $r$ denote the spherical coordinates in physical space. Note that the normalization coefficients in Eqs.~(\ref{eq:vectmult}) and (\ref{eq:scmult}) differ from those used in other works (see for example \cite{barton1988internal,barton1989theoretical}).

The PWA of the vector multipoles is given by \cite{devaney1974multipole}
\bse
\begin{align}
\bs \Lambda_{l,m}^{(\text{I})}(\textbf{r})=&\int_{4\pi} \textbf{Z}_{l,m}(\textbf{u})e^{\im k\bt u \cdot \bt r} \ud \Omega,  \\
\bs \Lambda_{l,m}^{(\text{II})}(\textbf{r})=&\int_{4\pi} \textbf{Y}_{l,m}(\textbf{u})e^{\im k\bt u \cdot \bt r} \ud\Omega,
\end{align}
\ese
where
\begin{subequations}
\begin{align}
\textbf{Z}_{l,m}(\theta,\phi)=&\textbf{u}\times \textbf{Y}_{l,m}(\theta,\phi), \\
\textbf{Y}_{l,m}(\theta,\phi)=&\frac{1}{\sqrt{l(l+1)}} \textbf{L}_\textbf{u} Y_{l,m}(\theta, \phi),
\end{align}
\end{subequations}
are the vector spherical harmonics and
\begin{align}
\textbf{L}_\textbf{u}=-\im \textbf{u}\times \nabla_\Omega=\im \frac{\hat{\bs \theta}}{\sin \theta}\pd{}{\phi} - \im \hat{\bs \phi}\pd{}{\theta},
\end{align}
is the angular momentum operator in PWA space. This is analogous to the relationship satisfied in the scalar case:
\begin{align}
 \Lambda_{l,m}(\textbf{r})=&\int_{4\pi} Y_{l,m}(\textbf{u})e^{\im k\bt u \cdot \bt r} \ud \Omega.
\end{align}
The vector multipoles form a complete orthonormal basis for monochromatic electromagnetic free fields \cite{sheppard1997efficient,jackson1998classical,
nieminen2003multipole,borghi2005highly}, and are central to Lorenz-Mie scattering theory since they constitute the appropriate set to satisfy the boundary conditions when the scatterer is spherical \cite{barton1988internal,barton1989theoretical}. 

\section{Bases of CF fields}

\subsection{Complete bases}

As mentioned in the Introduction, the incident fields used for trapping and manipulation must be highly focused in order to produce the necessary intensity gradient to counteract radiation pressure. However, most approaches to describe this type of field involve integral expressions  rather than closed forms. This is not the case for the CF constructions proposed in  \cite{moore2009closed,gutierrez-cuevas2017scalar}, which are given by analytic expressions and can incorporate properties of interest such as OAM, different states of polarization and a controllable degree of focusing \cite{dholakia2011shaping}.

\begin{figure}
\centering
\includegraphics[width=.99\linewidth]{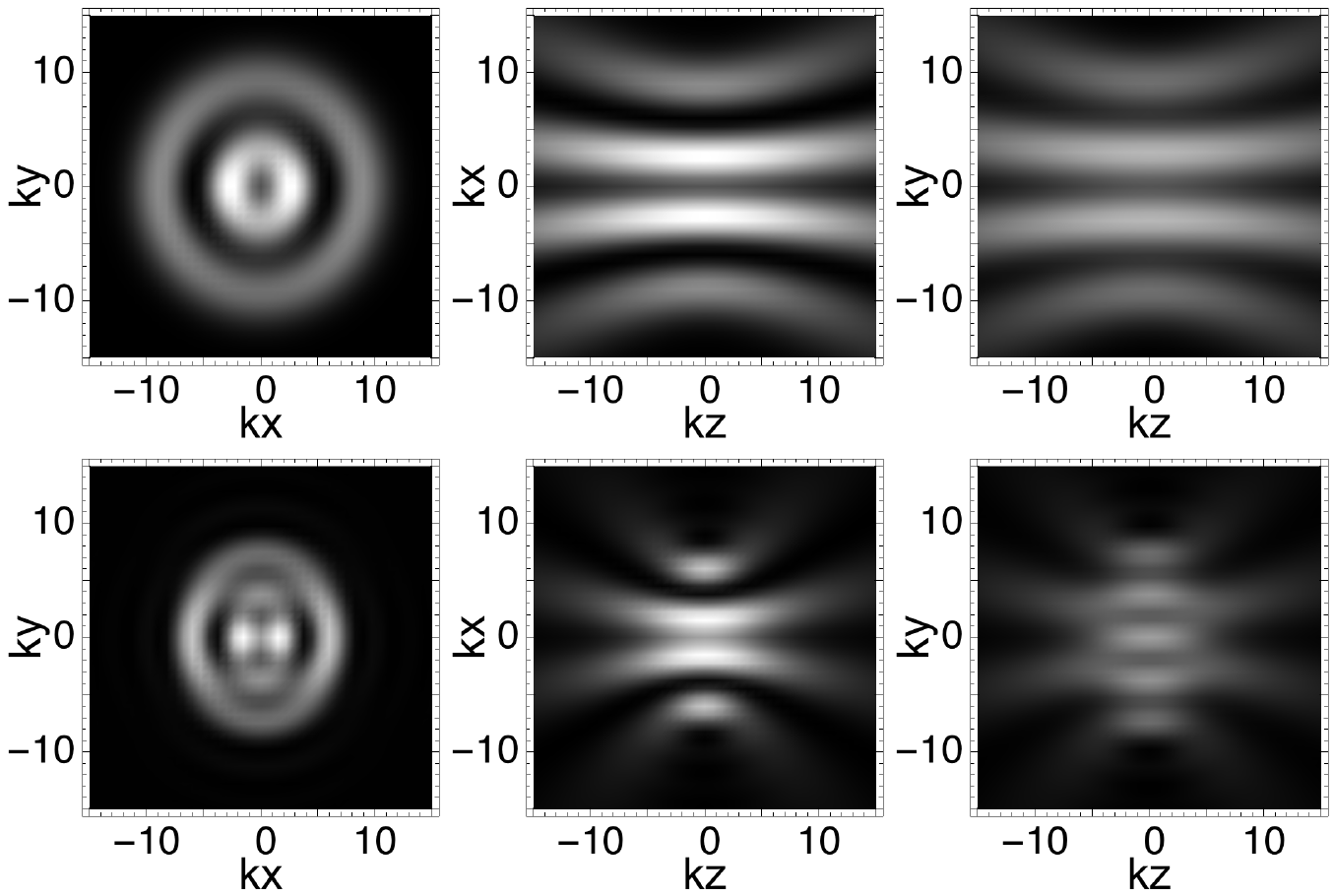}
 \caption{\label{fig:fieldLG11} Intensity profiles over the Cartesian planes for the basis element $\boldsymbol{\mathcal{L}}^{(\text{II})}_{1,1}$ quasi-linearly polarized along $y$ with $kq=15$ ($5$) in the first (second) row.}
\end{figure}

The main idea behind CF fields can be 
understood through the shift-phase property of Fourier transforms. Consider any free field that is described by a closed form expression. 
Clearly, the field that results from a spatial displacement of this expression is also expressible in closed form, even if such a 
displacement is complex.
Since the PWA space is a reduced version of Fourier space, 
a spatial shift by an imaginary distance in the $z$ direction ($z\rightarrow z-iq$) amounts to multiplication of the PWA by a real 
exponential of the form $\exp(kq \cos \theta)$. This factor weights more heavily the directions around the positive $z$ direction (namely 
$\theta =0$) thus inducing a controllable degree of directionality through the parameter $q$. 
Figure \ref{fig:fieldLG11} illustrates this effect for one of the basis elements considered here: as $kq$ decreases, the field becomes 
more focused, and nonparaxial effects become more apparent, such as the loss of rotational symmetry of the focal spot.

Different sets of complete bases can be constructed through complex displacements of multipoles. 
Here, we consider the two options proposed in Refs.~\cite{moore2009closed,gutierrez-cuevas2017scalar}. For the first 
\cite{moore2009closed}, the elements are given by weighted superpositions of complex multipoles with equal imaginary 
displacement and different radial indices:
\bse
\label{eq:lginmult}
\begin{align}
\boldsymbol{\mathcal{L}}^{(\text{I})}_{n,m}(\textbf{r};q) 
=& \frac{1}{\im k}\sum_{p=0}^n \alpha_{n,m}^{(p)}(q) \nabla \!\! \times \! \left[ \textbf{V}_\textbf{r}^\dagger \Lambda_{|m|+p,m}(\textbf{r}-\im q\hat{\textbf{z}})\right],\\
\boldsymbol{\mathcal{L}}^{(\text{II})}_{n,m}(\textbf{r};q) 
=& \sum_{p=0}^n \alpha_{n,m}^{(p)}(q)\textbf{V}_\textbf{r}\Lambda_{|m|+p,m}(\textbf{r}-\im q\hat{\textbf{z}}),
\end{align}
\ese
where $\textbf{V}_\textbf{r}$ is a polarization operator to be discussed in the next section. 
The coefficients $\alpha_{n,m}^{(p)}(q)$ depend on the choice of polarization and can be calculated using the expressions provided in 
Appendix \ref{app:aNb}. In the paraxial regime ($kq\gg 1$), the 
elements of this basis reduce to the standard LG beams; we therefore refer to them here as the nonparaxial Laguerre-Gauss 
(NLG) fields. The second type of basis was presented in 
\cite{gutierrez-cuevas2017scalar}, and is given by a weighted sum of complex multipoles of equal indices but different imaginary 
displacements according to
\bse
\label{eq:pginmult}
\begin{align}
 \boldsymbol{\mathcal{Q}}^{(\text{I})}_{n,m}&(\textbf{r};q)
=\frac{1}{\im k}\sum_{p=0}^n \beta_{n,m}^{(p)}(q) \nabla \!\! \times \!\! \left\{\textbf{V}_\textbf{r}^\dagger\Lambda_{|m|,m}[\textbf{r}-\im(2p+1)q\hat{\textbf{z}}]\right\}, \\
 \boldsymbol{\mathcal{Q}}^{(\text{II})}_{n,m}&(\textbf{r};q) 
=\sum_{p=0}^n \beta_{n,m}^{(p)}(q) \textbf{V}_\textbf{r}\Lambda_{|m|,m}[\textbf{r}-\im(2p+1)q\hat{\textbf{z}}] .
\end{align}
\ese
Again, the coefficients $\beta_{n,m}^{(p)}(q)$ depend on the choice of polarization and can be calculated using the formulas given in the 
Appendix \ref{app:aNb}. Two variants of this type of basis are considered \cite{gutierrez-cuevas2017scalar}: one that is orthogonal 
in PWA space but that requires non-standard polynomials, and one that is not exactly orthogonal but that is expressible in terms of 
Jacobi polynomials. These correspond to nonparaxial extensions of the bases expressed as polynomials of Gaussians 
\cite{gutierrez-cuevas2017polynomials,
gutierrez-cuevas2017complete}; we then refer to them as nonparaxial polynomials-of-Gaussians (NPG) fields.

\begin{figure}
\centering
\includegraphics[width=.99\linewidth]{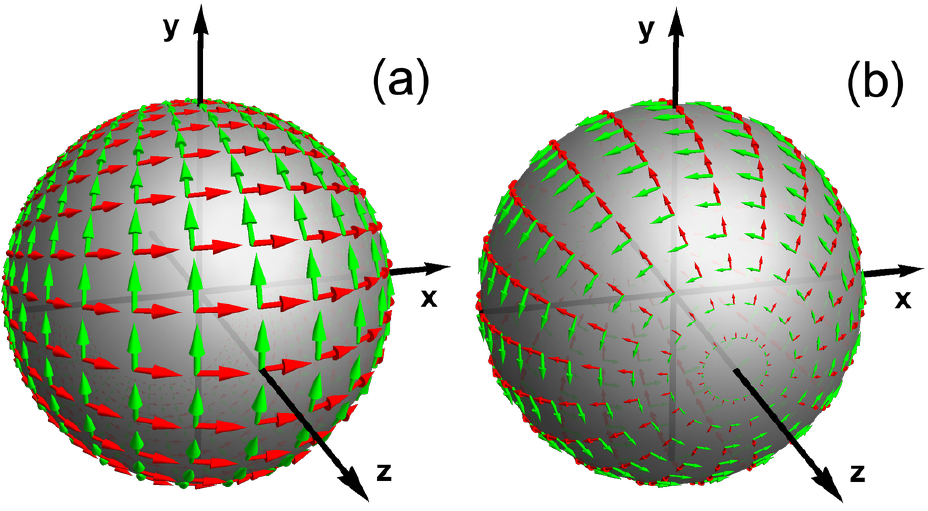}
\caption{\label{fig:polspheres} Polarization vectors,$\textbf{V}_\textbf{u}$ and $\bt u \times \textbf{V}_\textbf{u}^\dagger$,  in PWA space for the (a) quasi-linear and (b) TE-TM polarizations.}
\end{figure}

\subsection{Polarization operators}

The operators $\textbf{V}_\textbf{r}$ and $(\im k)^{-1} \nabla \! \times \! \textbf{V}_\textbf{r}^\dagger $ in Eqs.~(\ref{eq:lginmult}) 
and (\ref{eq:pginmult}) determine the two orthogonal polarization distributions employed by the elements of the bases. It is convenient 
to define them in terms of their PWA representation, which is related to that in physical space by the substitution 
$\bt u \leftrightarrow \nabla /\im k$. Hence, the first operator, written as  
$\textbf{V}_\textbf{u}$, must satisfy the transversality condition 
$\bt u \cdot \textbf{V}_\textbf{u}=0$; the second operator, given by 
$\bt u \times \textbf{V}_\textbf{u}^\dagger$, satisfies this condition 
automatically. The polarization  operators considered here are 
multiplicative in the PWA space while differential in physical space. 
(Other options are used e.g. in \cite{moore2009closed}.) 

The operators used here are of the general form
\bse
\label{eq:genV}
\begin{align}
\bt V_{\bt u}=&\textbf{V}_\textbf{u}^{(E)}(\bt p_1)+\textbf{V}_\textbf{u}^{(M)}(\bt p_2), \\
\bt u \times \bt V_{\bt u}^\dagger =&\textbf{V}_\textbf{u}^{(M)}(\bt p_1^* )-\textbf{V}_\textbf{u}^{(E)}(\bt p_2^* ),
\end{align}
\ese 
where the electric-like and magnetic-like dipolar distributions are defined as
\bse
\begin{align}
\textbf{V}_\textbf{u}^{(E)}(\bt p)=&\textbf{u}\times {\textbf{p}}\times \textbf{u}= \bt p -\bt u (\bt u \cdot \bt p), \\
\bt V_{\bt u}^{(M)}(\bt p)=&\bt u \times \textbf{V}_\textbf{u}^{(E)}(\bt p)= \bt u \times \bt p,
\end{align}
\ese
with $\textbf{p}$ being a vector indicating the direction of the dipole moment. 
Three particular polarization distributions of this type are considered, all of which resemble the focusing of a collimated field with a simple 
incident polarization according to the Richards-Wolf theory \cite{richards1959electromagnetic}. These are (i) the ``quasi-linear'' 
polarization basis, which resembles the focusing of linearly polarized beams in the $x$ and $y$ directions and for which 
$\textbf{V}_\textbf{u}=\textbf{V}_\textbf{u}^{(E)}(\hat{\textbf{x}})-\textbf{V}_\textbf{u}^{(M)}(\hat{\textbf{y}})$; 
(ii) the ``quasi-circular'' polarization basis, which resembles the 
focusing of beams with circular polarization $\bs \epsilon_\pm =(\hat{\textbf{x}}\pm \im \hat{\textbf{y}})/2$, for which 
$\textbf{V}_\textbf{u}=2^{1/2}\exp(\im\pi /4) [\textbf{V}_\textbf{u}^{(E)}(\bs \epsilon_+) + \im \textbf{V}_\textbf{u}^{(M)}(\bs \epsilon_+)]
$;
and (iii) the ``TE-TM'' polarization basis, which resembles the 
focusing of beams with azimuthal and radial polarizations, for which
$\textbf{V}_\textbf{u}=-\textbf{V}_\textbf{u}^{(E)}(\hat{\textbf{z}})$. Figure \ref{fig:polspheres} shows the polarization vectors along the 
sphere of directions corresponding to the quasi-linear and TE-TM operators.
It is worth mentioning that for the radial and azimuthal polarizations 
the paraxial limit leads not to LG and PG beams but to related cylindrical vector beams \cite{zhan2009cylindrical}.

\section{Scattering}

\subsection{Translation equation}

The first step for solving the scattering problem is to express the field in terms of vector multipoles centered around the scattering 
particle, assumed to be located at the origin (which does not 
necessarily coincide with the focus of the incident field).
To do this for the elements of both types of  basis, the decomposition of 
$\textbf{V}_\textbf{r}^{(E)}(\bt p)\Lambda_{l,m}(\textbf{r}-\bs \rho_0)$  and 
$\textbf{V}_\textbf{r}^{(M)}(\bt p)\Lambda_{l,m}(\textbf{r}-\bs \rho_0) $ 
is required, where $\bs \rho_0$ is a complex shift. This 
allows us to treat simultaneously the imaginary shift of the CF fields controlling the directionality through the parameter $q$ as well as 
the real shift, $\bt r_0$, locating the focus of the field with respect to 
the scatterer. It is shown in Appendix \ref{app:scattvec} that the result is
\begin{subequations}
\label{eq:transtomult}
\begin{align}
\textbf{V}_\textbf{r}^{(E)}(\bt p)&\Lambda_{l,m}(\textbf{r}-\bs \rho_0) \\
=&\sum_{L,M} \Big[ \eta_{L,M}^{(l,m)} 
\bs \Lambda_{L,M}^{(\text{I})}(\bt r) +
\xi_{L,M}^{(l,m)} \bs \Lambda_{L,M}^{(\text{II})}(\bt r) \Big]  ,\nonumber 
\end{align}
\begin{align}
\textbf{V}_\textbf{r}^{(M)}(\bt p)&\Lambda_{l,m}(\textbf{r}-\bs \rho_0) \\
=&\sum_{L,M} \Big[ \xi_{L,M}^{(l,m)}
\bs \Lambda_{L,M}^{(\text{I})}(\bt r) -
\eta_{L,M}^{(l,m)} \bs \Lambda_{L,M}^{(\text{II})}(\bt r) \Big], \nonumber
\end{align}
\end{subequations}
where
\begin{widetext}
\begin{subequations}
\label{eq:coefftrans}
\begin{align}
 \eta_{L,M}^{(l,m)}(\boldsymbol{\rho}_0,\bt p) =&\im \textbf{p} \cdot \Bigg\{ 
\sqrt{\frac{(L+1)}{L(2L+1)(2L-1)}} \Bigg[ 
-\sqrt{(L-M)(L-M-1)}\gamma_{L-1,M+1}^{(l,m)}(\boldsymbol{\rho}_0) \bs \epsilon_+ \nonumber \\
& + \sqrt{(L+M)(L+M-1)}\gamma_{L-1,M-1}^{(l,m)}(\boldsymbol{\rho}_0)\bs \epsilon_-
-\sqrt{(L-M)(L+M)}\gamma_{L-1,M}^{(l,m)}(\boldsymbol{\rho}_0)\hat{ \bt z}
\Bigg]\nonumber \\
&
+\sqrt{\frac{L}{(L+1)(2L+1)(2L+3)}}  \Bigg[ 
-\sqrt{(L+M+2)(L+M+1)}\gamma_{L+1,M+1}^{(l,m)}(\boldsymbol{\rho}_0)\bs \epsilon_+ \nonumber \\
&
+\sqrt{(L-M+2)(L-M+1)}\gamma_{L+1,M-1}^{(l,m)}(\boldsymbol{\rho}_0)\bs \epsilon_- 
+\sqrt{(L+M+1)(L-M+1)}\gamma_{L+1,M}^{(l,m)}(\boldsymbol{\rho}_0)\hat{ \bt z}
\Bigg] \Bigg\}, \\
\xi_{L,M}^{(l,m)}(\boldsymbol{\rho}_0,\bt p) =&\frac{1}{\sqrt{L(L+1)}}\bt p \cdot \Big[\sqrt{(L-M)(L+M+1)} \gamma_{L,M+1}^{(l,m)}(\boldsymbol{\rho}_0) \bs \epsilon_+ + \sqrt{(L+M)(L-M+1)}\gamma_{L,M-1}^{(l,m)}(\boldsymbol{\rho}_0)\bs \epsilon_- \nonumber \\
&+ M \gamma_{L,M}^{(l,m)}(\boldsymbol{\rho}_0)\hat{\textbf{z}} 
\Big].
\end{align}
\end{subequations}
\end{widetext}
Here, $\gamma_{l',m'}^{(l,m)}$ are the coefficients of the corresponding scalar translation equation \cite{moore2016mie},
\begin{align}
\label{eq:lambdashift}
\Lambda_{l,m}(\textbf{r}-\boldsymbol{\rho}_0)=\sum_{l',m'} \gamma_{l',m'}^{(l,m)}(\boldsymbol{\rho}_0) \Lambda_{l',m'}(\textbf{r}),
\end{align}
where 
\begin{align} 
\label{eq:gamma}
\gamma_{l',m'}^{(l,m)}(\boldsymbol{\rho}_0)=&\int Y_{l',m'}^*(\bt u)Y_{l,m}(\bt u) e^{-\im k \bs \rho_0 \cdot \textbf{u}} d \Omega  \nonumber \\
=&\sum_{j=|l-l'|}^{l+l'}(-1)^{m'}\sqrt{\frac{(2l'+1)(2j+1)(2l+1)}{4\pi }} \nonumber\\
&\times \left( \begin{array}{ccc}
l'&j&l \\
0&0&0
\end{array}\right)
\left( \begin{array}{ccc}
l'&j&l \\
-m'&m'-m&m
\end{array}\right) \nonumber \\
&\times \Lambda_{j,m'-m}^*(\boldsymbol{\rho}_0^*),
\end{align}
with $\left( \begin{array}{ccc}
l_1&l_2&l_3 \\
m_1&m_2&m_3
\end{array}\right) $ being the Wigner $3j$ symbols. Note that these coefficients differ from zero only if $j$ has the same parity as $l+l'$, so the sum in Eq.~(\ref{eq:gamma}) is in steps of two.

Using this translation equation and the general form for the polarization operators [Eq.~(\ref{eq:genV})], the translated elements of the bases can be written in terms of centered vector multipoles
(we write $\bs{\mathcal U}^{(\text{I},\text{II})}_{n,m}$ to denote either $\boldsymbol{\mathcal{L}}^{(\text{I},\text{II})}_{n,m}$ or $\boldsymbol{\mathcal{Q}}^{(\text{I},\text{II})}_{n,m}$) as
\bse
\label{eq:multexp}
\begin{align}
\bs{\mathcal U}^{(\text{I})}_{n,m}(\textbf{r}-\bt r_0;q)=\sum_{L,M} \Big( \bar{\mu}^{(n,m)}_{L,M}\bs \Lambda_{L,M}^{(\text{I})}(\bt r) -\bar{\upsilon}^{(n,m)}_{L,M} \bs \Lambda_{L,M}^{(\text{II})}(\bt r)  \Big),  \\
\bs{\mathcal U}^{(\text{II})}_{n,m}(\textbf{r}-\bt r_0;q)=\sum_{L,M} \Big( \upsilon^{(n,m)}_{L,M} \bs \Lambda_{L,M}^{(\text{I})}(\bt r) +\mu^{(n,m)}_{L,M} \bs \Lambda_{L,M}^{(\text{II})}(\bt r)  \Big) .
\end{align}
\ese
The complete dependence of the coefficients and exact form is given as follows: for the NLG basis,
\bse
\begin{align}
\prescript{}{\text{LG}}\upsilon^{(n,m)}_{L,M}&(\text r_0,q; \bt p_1,\bt p_2) \nonumber \\
=&\sum_{p=0}^n \alpha_{n,m}^{(p)}(q)  
\Big[ \eta_{L,M}^{(|m|+p,m)}(\bt r_0+\im q\hat{\textbf{z}},\bt p_1) \nonumber \\
&+\xi_{L,M}^{(|m|+p,m)}(\bt r_0+\im q\hat{\textbf{z}},\bt p_2) \Big] , \\
\prescript{}{\text{LG}}\mu^{(n,m)}_{L,M}&(\text r_0,q; \bt p_1,\bt p_2) \nonumber \\
=&\sum_{p=0}^n \alpha_{n,m}^{(p)}(q)  
\Big[ \xi_{L,M}^{(|m|+p,m)}(\bt r_0+\im q\hat{\textbf{z}},\bt p_1) \nonumber \\
&-\eta_{L,M}^{(|m|+p,m)}(\bt r_0+\im q\hat{\textbf{z}},\bt p_2) \Big],
\end{align}
\ese
and for the NPG basis,
\bse
\begin{align}
\prescript{}{\text{PG}}\upsilon^{(n,m)}_{L,M}&(\text r_0,q; \bt p_1,\bt p_2) \nonumber \\
=&\sum_{p=0}^n \beta_{n,m}^{(p)}(q)  
\Big\{ \eta_{L,M}^{(|m|,m)}[\bt r_0+\im (2p+1)q\hat{\textbf{z}},\bt p_1]  \nonumber \\
&+\xi_{L,M}^{(|m|,m)}[\bt r_0+\im (2p+1)q\hat{\textbf{z}},\bt p_2] \Big\},
\\
\prescript{}{\text{PG}}\mu^{(n,m)}_{L,M}&(\text r_0,q; \bt p_1,\bt p_2) \nonumber \\
=&\sum_{p=0}^n \beta_{n,m}^{(p)}(q)  
\Big\{\xi_{L,M}^{(|m|,m)}[\bt r_0+\im (2p+1)q\hat{\textbf{z}},\bt p_1]\nonumber \\
&-\eta_{L,M}^{(|m|,m)}[\bt r_0+\im (2p+1)q\hat{\textbf{z}},\bt p_2] \Big\}.
\end{align}
\ese
For the sake of brevity, we use the shorthand $\bar{\mu}^{(n,m)}_{L,M}=\mu^{(n,m)}_{L,M}(\text r_0,q; \bt p_1^*,\bt p_2^*)$ (and similarly for $\upsilon^{(n,m)}_{L,M}$) and omit the dependence in what follows.

\subsection{Scattering of basis elements}

When the incident field is any of the elements of the NLG or NPG bases, by virtue of the multipole expansion derived in the previous section, it can be written as
\begin{align}
\textbf{E}^{(\text{i})}(\textbf{r}-\textbf{r}_0)=E_0\!\!\sum_{L,M}\left[\kappa^{(\text{I})}_{L,M}\bs\Lambda^{(\text{I})}_{L,M}(\textbf{r}) +\kappa^{(\text{II})}_{L,M}\bs\Lambda^{(\text{II})}_{L,M}(\textbf{r}) \right],
\end{align}
where $\kappa^{(\text{I,II})}_{L,M}$ are the appropriate coefficients from Eqs.~(\ref{eq:multexp}) and $E_0$ is a constant amplitude factor. The scattered field, $\textbf{E}^{(\text{s})}$, must be expressed in terms of outgoing vector multipoles, $\bs \Pi^{(\text{I})}_{l,m}$ and $\bs \Pi^{(\text{II})}_{l,m}$, which have the same form as the regularized vector multipoles in Eq.~(\ref{eq:vectmult}), with the exception of the replacement of the spherical Bessel function, $j_l$, by the spherical Hankel functions of the first kind, $h^{(1)}_l$, thus giving
\begin{align}
\textbf{E}^{(\text{s})}(\textbf{r})=E_0 \!\!\sum_{L,M}\left[\chi^{(\text{I})}_{L,M}\bs\Pi^{(\text{I})}_{L,M}(\textbf{r}) +\chi^{(\text{II})}_{L,M}\bs\Pi^{(\text{II})}_{L,M}(\textbf{r}) \right].
\end{align}
Since the scattering process is linear, the coefficients of the scattered field are related to those of the incident field via
\begin{align}
\bs \chi = \bt T \cdot \bs \kappa,
\end{align}
where $\bs \kappa$ ($\bs \chi$) is the ordered vector of the incident (scattered) field coefficients, $\kappa^{(\text{I,II})}_{L,M}$ [$\chi^{(\text{I,II})}_{L,M}$], and $\bt T$ is a matrix (conveniently called the T-matrix) that incorporates all the relevant information of the scattering particle \cite{mishchenko1999light,nieminen2011t,gouesbet2017generalized}.

\begin{figure}
\centering
\includegraphics[width=.99\linewidth]{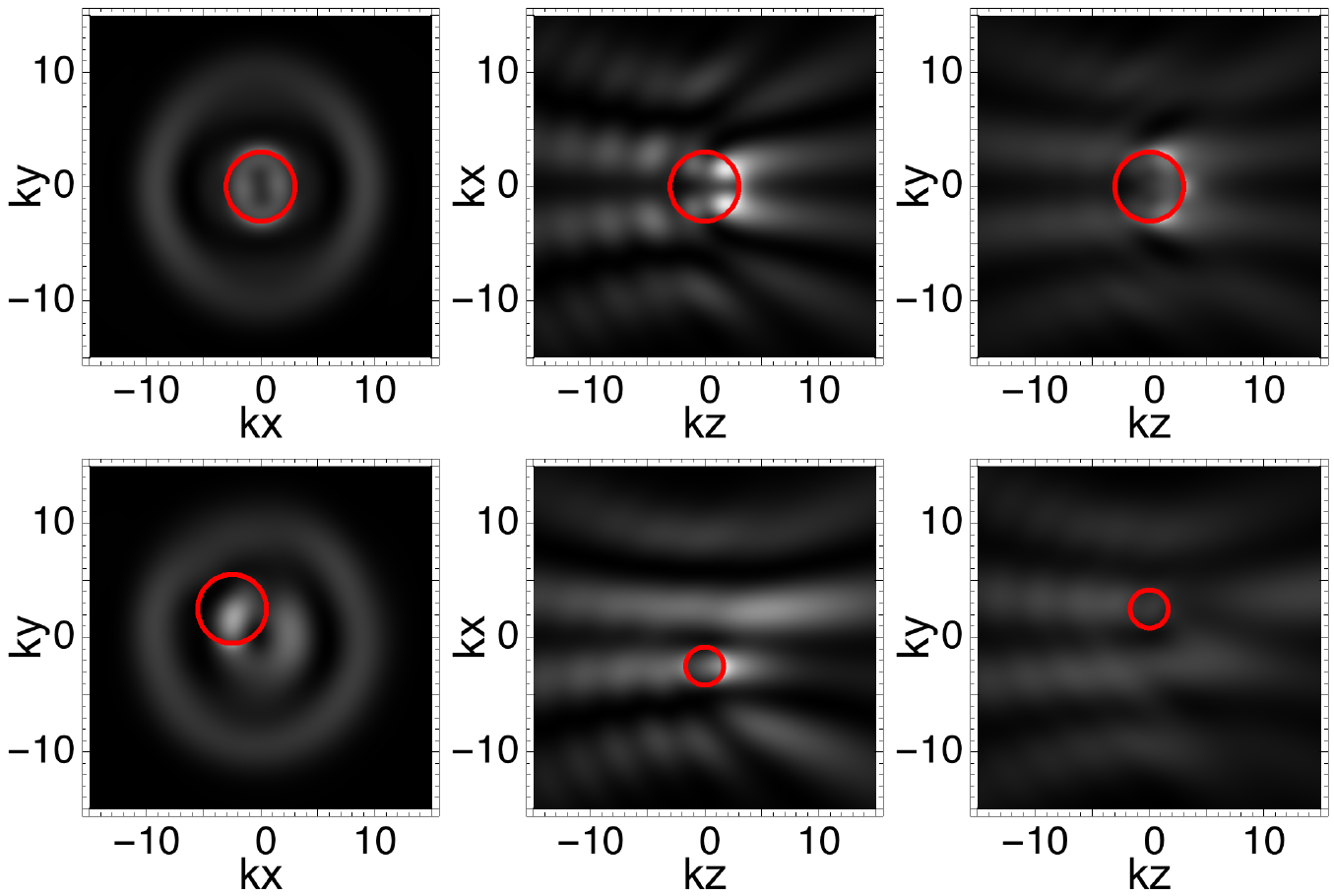}
\caption{\label{fig:scattLG11} Intensity profiles over the Cartesian planes for the total field generated by the scattering of the basis element $\boldsymbol{\mathcal{L}}^{(\text{II})}_{1,1}$ quasi-linearly-polarized along $y$ with $kq=15$ by a spherical particle of relative index of refraction $\nu_0=1.38+10^{-4} \im$ and radius $kR=3$ located at $\bt r_p=(0,0,0)$ [$(-2.5,2.5,0)$] in the first (second) row. Note that the position of the particle is specified relative to the focus of the beam and the red circles show the cross-section of the sphere by the corresponding plane 
(hence the smaller size in the last two figures).}
\end{figure}

For the particular case of a spherical scatterer of radius $R$ and relative (to the external medium) index of refraction $\nu_0$, the T-matrix is diagonal and its entries are the well-known Mie coefficients \cite{barton1988internal},
\bse
\begin{align}
a^{(\text{s})}_l(kR,\nu_0) =& \frac{ \psi_l(kR)\psi_l'(k\nu_0R)-\nu_0\psi_l(k\nu_0R)\psi_l'(kR)}{\nu_0\psi_l(k\nu_0R)\zeta_l'(kR)- \zeta_l(kR)\psi_l'(k\nu_0R)}, \\
b^{(\text{s})}_l(kR,\nu_0) =& \frac{\nu_0 \psi_l(kR)\psi_l'(k\nu_0R)-\psi_l(k\nu_0R)\psi_l'(kR)}
{\psi_l(k\nu_0R)\zeta_l{}'(kR)-\nu_0 \zeta_l(kR)\psi_l'(k\nu_0R)},
\end{align}
\ese
with $\zeta_l(z)=z h_l^{(1)}(z)$ and $\psi_l(z)=z j_l(z)$.
The corresponding scattered field can then be written as
\begin{align}
\textbf{E}^{(\text{s})}(\textbf{r})=E_0\!\!\sum_{L,M}\left[a^{(\text{s})}_{L}\kappa^{(\text{I})}_{L,M}\bs\Pi^{(\text{I})}_{L,M}(\textbf{r}) +b^{(\text{s})}_{L}\kappa^{(\text{II})}_{L,M}\bs\Pi^{(\text{II})}_{L,M}(\textbf{r}) \right].
\end{align}
A similar expression can be obtained for the internal field, $\textbf{E}^{(\text{w})}$, although expressed in terms of regularized vector multipoles:
\begin{align}
\textbf{E}^{(\text{w})}(\textbf{r})=&E_0\!\!\sum_{L,M}\left[a^{(\text{w})}_{L}\kappa^{(\text{I})}_{L,M}\bs\Lambda^{(\text{I})}_{L,M}(\nu_0\textbf{r}) \right. \nonumber \\ &\left.+b^{(\text{w})}_{L}\kappa^{(\text{II})}_{L,M}\bs\Lambda^{(\text{II})}_{L,M}(\nu_0\textbf{r}) \right],
\end{align}
where
\bse
\begin{align}
a^{(\text{w})}_l(kR,\nu_0) =& \frac{ \nu_0\psi_l(kR)\zeta_l'(kR)-\nu_0\zeta_l(kR)\psi_l'(kR)}{\nu_0\psi_l(k\nu_0R)\zeta_l'(kR)- \zeta_l(kR)\psi_l'(k\nu_0R)}, \\
b^{(\text{w})}_l(kR,\nu_0) =& \frac{\nu_0 \psi_l(kR)\zeta_l'(kR)-\nu_0\zeta_l(kR)\psi_l'(kR)}
{\psi_l(k\nu_0R)\zeta_l{}'(kR)-\nu_0 \zeta_l(kR)\psi_l'(k\nu_0R)}.
\end{align}
\ese
Figure \ref{fig:scattLG11} shows the total field after the scattering of the quasi-linearly-polarized field $\boldsymbol{\mathcal{L}}^{(\text{II})}_{1,1}$ with $kq=15$ by a spherical particle of radius $kR=3$ and relative index of refraction $\nu_0=1.38+10^{-4}\im$. Note that the plots are in the field's system of reference, $\bt r_f=\bt r -\bt r_0$, in which the particle location is simply given by $\bt r_p =-\bt r_0$.

\begin{figure}
\centering
\includegraphics[width=.99\linewidth]{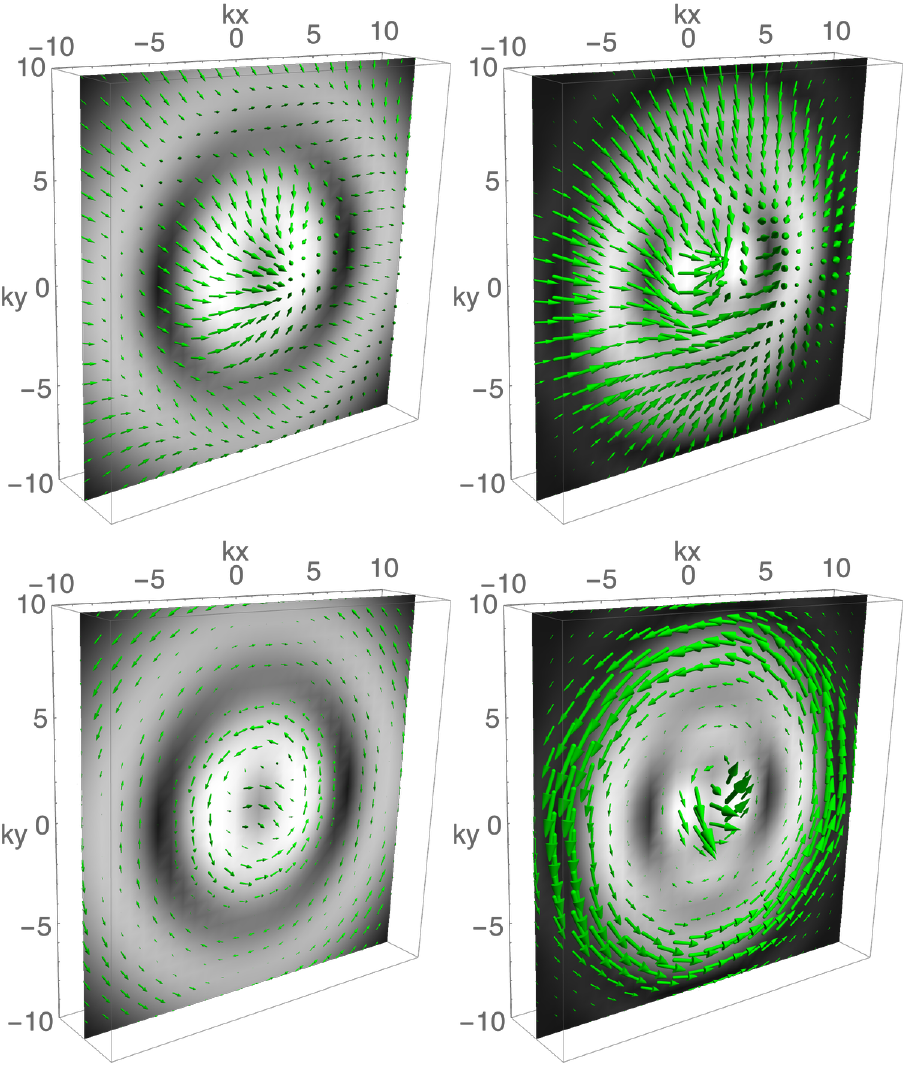}
\caption{\label{fig:fNtLG11} Force (first row) and torque (second row) field maps overlaid on the corresponding intensity along the $x-y$ plane. The incident field  is the quasi-linearly-polarized element $\boldsymbol{\mathcal{L}}^{(\text{II})}_{1,1}$  with $kq=15$ ($5$) in the first (second) column and the scattering particle is the same as in Fig.~\ref{fig:scattLG11}.}
\end{figure}

Having the total field, the forces and torques exerted on the spherical scatterer can be computed at any given point. These are obtained after integration of the force and torque densities expressed in terms of Maxwell's stress tensor. Adapting the results presented in \cite{barton1989theoretical} to our notation gives
\begin{widetext}
\bse
\begin{align}
F_x+\im F_y=&\frac{4\pi^2 E_0^2\varepsilon \im}{k^2} \sum_{L,M}\Bigg( \frac{-\im}{L+1}\sqrt{\frac{(L+M+2)(L+M+1)L(L+2)}{(2L+1)(2L+3)}}  
\Big\{ \Big[2a^{(\text{s})}_La^{(\text{s})*}_{L+1} +a_L+a^{(\text{s})*}_{L+1}\Big]\kappa^{(\text{I})}_{L,M}\kappa^{(\text{I})}_{L+1,M+1}{}^* \nonumber \\
&+\Big[2b^{(\text{s})}_Lb^{(\text{s})*}_{L+1} +b^{(\text{s})}_L+b^{(\text{s})*}_{L+1}\Big]\kappa^{(\text{II})}_{L,M}\kappa^{(\text{II})}_{L+1,M+1}{}^*\Big\} 
+\frac{\im}{L+1}\sqrt{\frac{(L-M+1)(L-M+2)L(L+2)}{(2L+1)(2L+3)}} \nonumber \\
& \times
\Big\{ \Big[ 2a_L^{(\text{s})*}a^{(\text{s})}_{L+1} +a_L^{(\text{s})*}+a^{(\text{s})}_{L+1}\Big]\kappa^{(\text{I})}_{L,M}{}^*\kappa^{(\text{I})}_{L+1,M-1}
+\Big[ 2b_L^{(\text{s})*}b^{(\text{s})}_{L+1} +b_L^{(\text{s})*}+b^{(\text{s})}_{L+1} \Big] \kappa^{(\text{II})}_{L,M}{}^*\kappa^{(\text{II})}_{L+1,M-1}\Big\}  \\
&
-\frac{\sqrt{(L+M+1)(L-M)}}{L(L+1)} \Big\{ \Big[ 2a_L^{(\text{s})*}b^{(\text{s})}_L+a_L^{(\text{s})*}+b_L^{(\text{s})}\Big] \kappa^{(\text{I})}_{L,M+1}{}^*\kappa^{(\text{II})}_{L,M}-\Big[2a_L^{(\text{s})}b_L^{(\text{s})*}+a_L^{(\text{s})}+b_L^{(\text{s})*}\Big] \kappa^{(\text{I})}_{L,M}\kappa^{(\text{II})}_{L,M+1}{}^*  \Big\}
\Bigg), \nonumber \\
F_z=& -\frac{8\pi^2E_0^2 \varepsilon }{k^2}\sum_{L,M} \text{Im} \Bigg( \frac{\im}{L+1} \sqrt{\frac{(L-M+1)(L+M+1)L(L+2)}{(2L+1)(2L+3)}}  \Big\{ \Big[ 2a_{L+1}^{(\text{s})}a_L^{(\text{s})*}+a_{L+1}^{(\text{s})}+a_L^{(\text{s})*} \Big] \kappa^{(\text{I})}_{L,M}{}^*\kappa^{(\text{I})}_{L+1,M}\nonumber \\
& +\Big[ 2b_{L+1}^{(\text{s})}b_L^{(\text{s})*}+b_{L+1}^{(\text{s})}+b_L^{(\text{s})*} \Big] \kappa^{(\text{II})}_{L,M}{}^*\kappa^{(\text{II})}_{L+1,M}  \Big\} 
+\frac{M}{L(L+1)}\Big[ 2a_L^{(\text{s})}b_L^{(\text{s})*}+a_L^{(\text{s})}+b_L^{(\text{s})*}\Big] \kappa^{(\text{I})}_{L,M}\kappa^{(\text{II})}_{L,M} {}^* \Bigg),
\end{align}
\ese
\end{widetext}
and
\bse 
\begin{align}
N_x+\im N_y=& -\frac{8\pi^2 E_0^2\varepsilon }{k^3} \sum_{L,M}  \sqrt{(L-M)(L+M+1)} \nonumber \\
&\times \Bigg( \Big\{ |a_L^{(\text{s})}|^2 + \text{Re} \Big[ a_L^{(\text{s})}\Big] \Big\} \kappa^{(\text{I})}_{L,M}\kappa^{(\text{I})}_{L,M+1}{}^* \nonumber \\
&+ \Big\{ |b_L^{(\text{s})}|^2 +\text{Re} \Big[ b_L^{(\text{s})} \Big] \Big\} \kappa^{(\text{II})}_{L,M}\kappa^{(\text{II})}_{L,M+1}{}^*  \Bigg), 
\end{align}
\begin{align}
N_z=& -\frac{8\pi^2 E_0^2\varepsilon }{k^3} \sum_{L,M} \frac{M}{L} \Big\{
|a_L^{(\text{s})} \kappa^{(\text{I})}_{L,M}|^2  + |b_L^{(\text{s})} \kappa^{(\text{II})}_{L,M}|^2 \nonumber \\
&+\text{Re}\Big[ a_L^{(\text{s})} | \kappa^{(\text{I})}_{L,M}|^2  + b_L^{(\text{s})} |\kappa^{(\text{II})}_{L,M}|^2 \Big]
\Big\}
\end{align}
\ese 
with $\varepsilon$ being the dielectric constant of the embedding medium.

As a simple example, Fig.~\ref{fig:fNtLG11} shows the force and torque field maps for the incident fields shown in Fig.~\ref{fig:fieldLG11} and the same scattering particle used for Fig.~\ref{fig:scattLG11}. Figure \ref{fig:fNtLG11} shows the effects of focusing on the forces and torques exerted on the scattering particle. 
Clearly, stronger focusing resutls in increased forces and torques, leading to enhanced trapping and manipulation capabilities.

\section{NLG vs NPG}

\begin{figure}
\centering
\includegraphics[width=.99\linewidth]{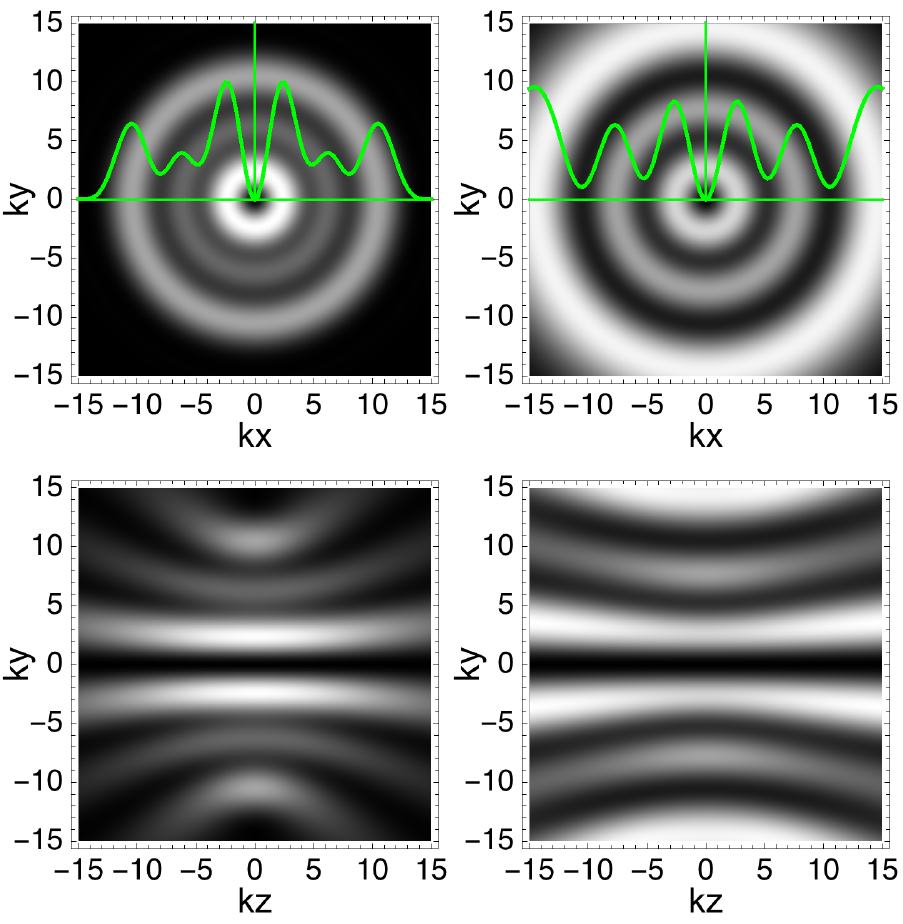}
\caption{\label{fig:field21} Intensity profiles over the Cartesian planes for the basis elements $\boldsymbol{\mathcal{L}}^{(\text{II})}_{2,1}$ (first column) and the orthogonal $\boldsymbol{\mathcal{Q}}^{(\text{II})}_{2,1}$ (second column) quasi-circularly polarized with $kq=10$. 
}
\end{figure}

We now provide a brief comparison between the trapping properties of elements of both the NLG and orthogonal NPG bases with similar radial structure, OAM, and imaginary displacement. Since their elements actually coincide for $n=0$, we use elements with higher-order radial structure. Consider the elements of both bases with $n=2$, $m=1$ and $kq=10$, shown in Fig.~\ref{fig:field21}. 
While their intensity distributions have common features, there are clear differences, e.g. the NLG basis is noticeably more focused even though the same $q$ was used for both. 

\begin{figure}
\centering
\includegraphics[width=1.\linewidth]{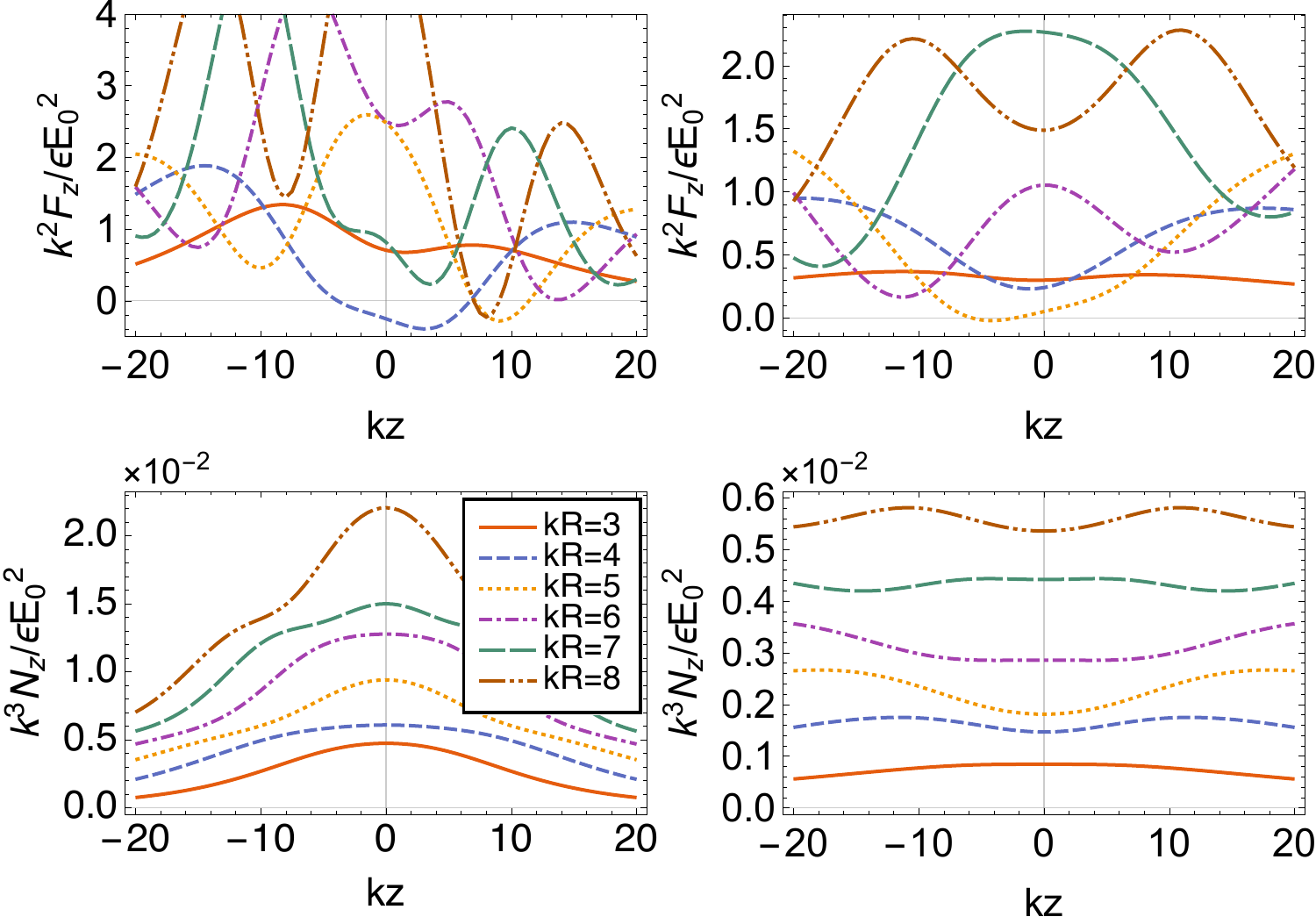}
\caption{\label{fig:fNtz21} Dimensionless force (first row) and torque (second row) along the field axis, exerted on different-sized spherical scatterers by the basis elements $\boldsymbol{\mathcal{L}}^{(\text{II})}_{2,1}$ (first column) and the orthogonal $\boldsymbol{\mathcal{Q}}^{(\text{II})}_{2,1}$ (second column) with quasi-circular polarization and $kq=10$. }
\end{figure}

\begin{figure}
\centering
\includegraphics[width=1.\linewidth]{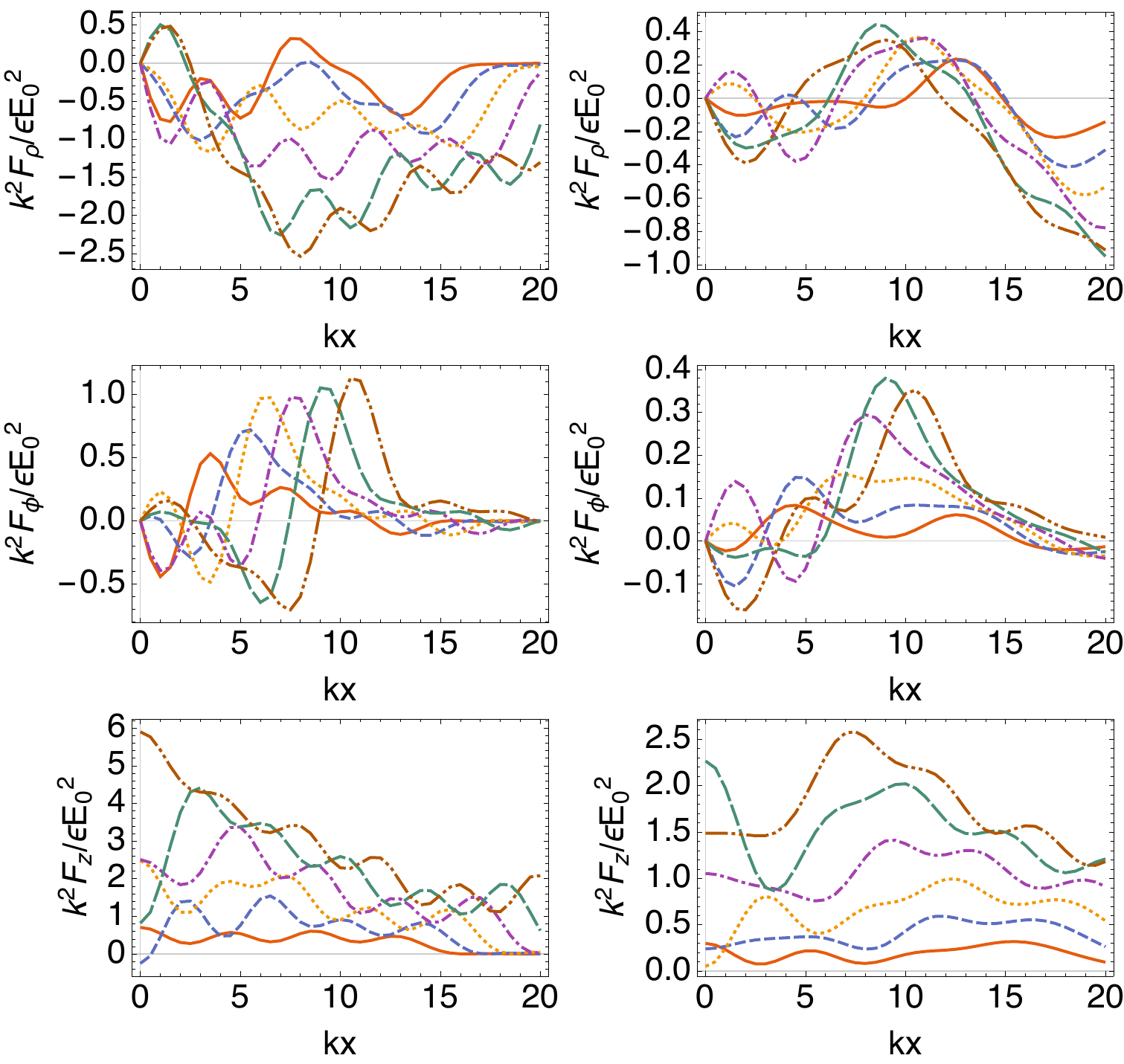}
\caption{\label{fig:radfNt21} Dimensionless radial  (first row), azimuthal  (second row) and axial  (third row) force along a transverse direction, exerted on different sized spherical scatterers by the basis elements $\boldsymbol{\mathcal{L}}^{(\text{II})}_{2,1}$ (first column) and the orthogonal $\boldsymbol{\mathcal{Q}}^{(\text{II})}_{2,1}$ (second column) with quasi-circular polarization and $kq=10$. The legend is the same as that in Fig.~\ref{fig:fNtz21}. }
\end{figure}

Figure \ref{fig:fNtz21} shows the dimensionless forces ($k^2F/\epsilon E_0^2$) and torques ($k^3N/\epsilon E_0^2$) exerted by these two fields 
on spherical scatterers of relative index of refraction $\nu_0=1.38+10^{-4}\im$ and different radii. 
A clear difference can be noticed in the axial force distribution of these fields, including their trapping capabilities: 
the NLG field can trap particles of radii $kR=4$, $5$ and $8$ since the corresponding curve has a stable equilibrium point (a zero of negative slope), whereas the NPG field could only trap a particle of size $kR=5$ (being optimistic). 
Another interesting difference is the location at which the maximum torque is achieved: for the NLG field this maximum is always located at the focus, while for the NPG field this location depends on the size of the particle, probably because the focal plane is not the location of maximum intensity for all the rings. 
The force components along a transverse direction at the focal plane are plotted in Fig.~\ref{fig:radfNt21}. 
An interesting feature that appears for both fields is the reversal of azimuthal force with respect to the vortex charge, although, this effect is less marked for the NPG field. This behavior has already been reported in multi-ringed fields \cite{volke-sepulveda2004three}.

\section{Concluding Remarks}

In summary, we presented the generalized Lorenz-Mie scattering theory for a wide class of focused electromagnetic fields, which correspond to the elements of complete sets of bases that can be constructed from CF fields. They are given by simple closed-form expressions that allow an analytic multipolar expansion (necessary for the use of GLMT).
Furthermore, they  exhibit many interesting properties, such as OAM, different polarization distributions and controllable degree of focusing, thus providing an appealing alternative for future research in Lorenz-Mie scattering and optical manipulation.

While the two types of field considered in this work present similar radial and angular structure, they cannot be used interchangeably. Their different functional form in terms of CF fields 
leads to 
noticeable differences in the forces and torques they exert on a scatterer. 
The choice between the NLG and NPG bases should be made according to whether the paraxial behavior of the incident field is best modeled by LG or PG beams, respectively.

Since these fields are elements of complete bases, they can be superposed to describe arbitrary incident fields. 
This approach is justified particularly if the number of elements needed to accurately describe the incident field in question is considerably lower than that the number of standard multipoles used in the decomposition.  Let us stress, however, that the main value of the fields studied here lies in their similarity to fields of interest for trapping and manipulation experiments. 
This point will be developed further in subsequent publications.

\begin{acknowledgments}
This work was supported by NSF grant PHY-1507278.  R.G.-C.  acknowledges support of a  CONACYT fellowship and M.A.A. acknowledges support from  the Excellence Initiative of Aix-Marseille University- A*MIDEX, a French ``Investissements d'Avenir'' programme
\end{acknowledgments}

\appendix

\section{Computation of the coefficients $\bs \alpha$, $\bs \beta$}
\label{app:aNb}

As mentioned in the main body, the elements of the bases are composed of multipoles vectorized by a polarization operator and displaced to a complex location. The coefficients $\alpha$ and $\beta$ in Eqs.~(\ref{eq:lginmult}) and (\ref{eq:pginmult}) control the amount of each of these building blocks in order to obtain complete bases. The recipe  to calculate these coefficients is given here for these bases and the polarization operators discussed in the main body. For further details we refer the interested reader to \cite{moore2009closed,gutierrez-cuevas2017scalar}.

For the orthogonal bases, the formulas result from using the method of moments \cite{szego1967orthogonal}.
An orthogonal set of polynomials with respect to the weight function $w(x)$ in the interval $[x_1,x_2]$ satisfies 
\begin{align}
\int_{x_1}^{x_2}f_n(x)w_m(x)f_{n'}(x)dx=h_n^{(m)} \delta_{n,n'}
\end{align}
where
\begin{align}
f_n(x)= K_n^{(n,m)}x^n +...+(-1)^{n-j}K_j^{(n,m)}+...+(-1)^n K_0^{(n,m)}.
\end{align}
The index $m$ on the weight relates to the topological charge of the fields.
The coefficients $K_j^{(n,m)}$ and the normalization constant $h_n^{(m)}$ can be computed in terms of the moments,
\begin{align}
\label{eq:mom}
\mu_j^{(m)}=\int_{x_1}^{x_2} w_m(x)x^jdx,
\end{align}
by the formulas
\begin{align}
K_j^{(n,m)}=\left|
\begin{array}{ccccccc}
\mu_0^{(m)}&\mu_1^{(m)}& \cdots       &	\mu_{j-1}^{(m)}  &	\mu_{j+1}^{(m)}& \cdots	 &\mu_n^{(m)}\\
\mu_1^{(m)}&\mu_2^{(m)}& \cdots       &	\mu_{j}^{(m)}	   &	\mu_{j+2}^{(m)}& \cdots	     &\mu_{n+1}^{(m)}\\
\vdots& \vdots & \ddots   & \vdots 	   & \vdots    & \ddots & \vdots\\
\mu_{n-1}^{(m)}&\mu_{n}^{(m)}& \cdots & \mu_{n+j-2}^{(m)}& \mu_{n+j}^{(m)} & \cdots	 &\mu_{2n-1}^{(m)}
\end{array}
\right|
\end{align}
and
\begin{align}
h_n^{(m)}=K_n^{(n,m)}K_{n+1}^{(n+1,m)}.
\end{align}

We now give the details for each basis and polarization distribution.

\textbf{NLG basis.}
In this case, the coefficients are given by the solution to the following system of linear equations:
\begin{align}
\left[\begin{array}{c}
\bar K_{n,m}^{(0)}\\
\bar K_{n,m}^{(1)}\\
\vdots \\
\bar K_{n,m}^{(n)}
\end{array}\right] = 
\left[\begin{array}{ccc}
p_{0,m}^{(0)}&\cdots & p_{n,m}^{(0)}\\
p_{0,m}^{(1)}&\cdots & p_{n,m}^{(1)}\\
\vdots & \ddots & \vdots \\
p_{0,m}^{(n)}&\cdots & p_{n,m}^{(n)}
\end{array}\right]
\left[\begin{array}{c}
\alpha_{n,m}^{(0)}\\
\alpha_{n,m}^{(1)}\\
\vdots \\
\alpha_{n,m}^{(n)}
\end{array}\right],
\end{align}
where
\begin{align}
\bar K_{n,m}^{(i)}=(-1)^{n-i}\frac{K_i^{(n,m)}}{\sqrt{h_n^{(m)}}},
\end{align}
and $p_{n,m}^{(i)}$ is the coefficient of the $i^\text{th}$ power of the polynomial
\begin{align}
\sigma_m^m\sqrt{\frac{(2n +2|m|+1)n!}{2(n+2|m|)!}}\frac{P_{|m|+n}^{(|m|)}(x)}{(1-x^2)^{|m|/2}}.
\end{align}
Note that the matrix $p_{n,m}^{(i)}$ is upper-triangular, which simplifies the solution to the system of equations. Alternatively, the coefficients $\alpha$ can be determined via a recurrence relation \cite{moore2009bases}.
The functional form of the weight is fixed by the polarization basis.
\\
\emph{Quasi-linear and quasi-circular.} The interval of integration for Eq.~(\ref{eq:mom}) is $[-1,1]$ and 
\begin{align}
w_m(x)=(1+x)^2(1-x^2)^{|m|}e^{2kqx}.
\end{align}
\\
\emph{TE-TM (radial and azimuthal).} The interval of integration for Eq.~(\ref{eq:mom}) is $[-1,1]$ and 
\begin{align}
w_m(x)=(1-x^2)^{|m|+1}e^{2kqx}.
\end{align}

\textbf{Orthogonal NPG.}
In this case, the coefficients are given by the simpler formula
\begin{align}
\beta_{n,m}^{(i)}=(-1)^{n-i} c_m \sqrt{\frac{kq}{\pi h_n^{(m)}}}K_i^{(n,m)},
\end{align}
where
\begin{align}
c_m=\frac{2^{|m|}|m|!}{(-\sigma_m)^{|m|} } \sqrt{\frac{4\pi}{(2|m|+1)!}}.
\end{align}
Again, the weight depends on the topological charge $m$ and its functional form is fixed by the polarization basis.
\\
\emph{Quasi-linear and quasi-circular.} The interval of integration for Eq.~(\ref{eq:mom}) is $[\exp(-2kq),\exp(2kq)]$ and 
\begin{align}
w_m(x)=\left(1+\frac{\ln x}{2kq}\right)^2\left[1-\left(\frac{\ln x}{2kq}\right)^2\right]^{|m|}.
\end{align}
\\
\emph{TE-TM (radial and azimuthal).} The interval of integration for Eq.~(\ref{eq:mom}) is $[\exp(-2kq),\exp(2kq)]$ and 
\begin{align}
w_m(x)=\left[1-\left(\frac{\ln x}{2kq}\right)^2\right]^{|m|}.
\end{align}

\textbf{Nonorthogonal NPG.}
This basis is expressible in terms of Jacobi polynomials.
and the coefficients are simply given by
\begin{align}
\beta_{n,m}^{(i)}=&\sqrt{\frac{kqn!(2n+2|m|+3)(n+2|m|+2)!}{2^{2|m|+3}\pi \sinh(2kq)(n+|m|+s)!(n+|m|+t)!}} \nonumber \\
 & \times c_m p_{n,m}^{(i)},
\end{align}
where $p_{n,m}^{(i)}$ is the coefficient of the $i^\text{th}$ power of the polynomial
\begin{align}
P_n^{(|m|+s,|m|+t)}\left[ \frac{x-\cosh(2kq)}{\sinh(2kq)}\right].
\end{align}
The integers $s$ and $t$ are determined by the polarization distribution.
\\
\emph{Quasi-linear and quasi-circular:} $s=0$ and $t=2$.
\\
\emph{TE-TM (radial and azimuthal):} $s=t=1$.

\section{Derivation of the translation equation}
\label{app:scattvec}

Using the translation equation for the scalar multipoles we can write
\begin{align}
\label{eq:sveas}
\textbf{V}_\textbf{r}^{(E)}\Lambda_{l,m}(\textbf{r}-\bs \rho_0) = \sum_{l',m'} \gamma_{l',m'}^{(l,m)}(\boldsymbol{\rho}_0)\textbf{V}_\textbf{r} ^{(E)} \Lambda_{l',m'}(\textbf{r}) \nonumber \\
=\sum_{l',m'} \gamma_{l',m'}^{(l,m)}(\boldsymbol{\rho}_0)\int_{4\pi} e^{\im k\textbf{u} \cdot  \textbf{r}} \textbf{V}_\textbf{u}^{(E)} Y_{l',m'}(\bt u)  d\Omega, 
\end{align}
and similarly for $\textbf{V}_\textbf{r}^{(M)}$. Then we expand the integrand in terms of the vector spherical harmonics, and since $\bt V_{\bt u}^{(M)}=\bt u \times \textbf{V}_\textbf{u}^{(E)}$, we have
\bse
\label{eq:veinyz} 
\begin{align}
 \textbf{V}_\textbf{u}^{(E)} Y_{l',m'}(\bt u)&=\!\!\sum_{L,M} a_{L,M}^{(l',m')} \bt Z_{L,M}(\bt u)+b_{L,M}^{(l',m')} \bt Y_{L,M}(\bt u), \\
  \textbf{V}_\textbf{u}^{(M)} Y_{l',m'}(\bt u)&=\!\!\sum_{L,M} b_{L,M}^{(l',m')} \bt Z_{L,M}(\bt u)- a_{L,M}^{(l',m')}\bt Y_{L,M}(\bt u),
\end{align}
\ese
where the coefficients are given by
\bse
\begin{align}
a_{L,M}^{(l',m')} =&\int \bt Z_{L,M}^*(\bt u) \cdot \textbf{V}_\textbf{u}^{(E)} Y_{l',m'}(\bt u) d\Omega,  \\
b_{L,M}^{(l',m')} =&\int \bt Y_{L,M}^*(\bt u) \cdot \textbf{V}_\textbf{u}^{(E)} Y_{l',m'}(\bt u) d\Omega. 
\end{align}
\ese
Recalling that $\textbf{V}_\textbf{u}^{(E)}= \bt p -\bt u (\bt u \cdot \bt p)$ and using $\bt u \cdot \bt Z_{l,m}=\bt u \cdot \bt Y_{l,m}=0$, we get
\bse
\begin{align}
\label{eq:afirstint}
a_{L,M}^{(l',m')}  =&\textbf{p}  \cdot \int \bt u \times \bt Y_{L,M}^*(\bt u) Y_{l',m'}(\bt u) d\Omega,  \\
b_{L,M}^{(l',m')}  =&\textbf{p}\cdot \int \bt Y_{L,M}^*(\bt u)  Y_{l',m'}(\bt u) d\Omega. 
\end{align}
\ese
Using the following identity \cite{devaney1974multipole}:
\begin{align}
\textbf{Y}_{l,m}(\bt u)=&\frac{1}{\sqrt{l(l+1)}} [c_{l,m}^{(-)}Y_{l,m+1}(\bt u)\bs \epsilon_- \nonumber\\
&+c_{l,m}^{(+)}Y_{l,m-1}(\bt u)\bs \epsilon_+ + m Y_{l,m}(\bt u) \hat{\textbf{z}} ],
\end{align}
where 
\begin{align}
c_{l,m}^{(\pm)}=&\sqrt{(l\pm m)(l\mp m+1) }, \\
\bs \epsilon_\pm =&\frac{1}{2}(\hat{\textbf{x}}\pm \im\hat{\textbf{y}}) ,
\end{align}
we can compute the second integral using the orthogonality between the scalar spherical harmonics:
\begin{align}
\label{eq:bcoef}
b_{L,M}^{(l',m')} =&\frac{1}{\sqrt{L(L+1)}}\bt p \cdot [c_{L,M}^{(-)}\delta_{l',L}\delta_{m',M+1}\bs \epsilon_+ \nonumber\\
&+c_{L,M}^{(+)}\delta_{l',L}\delta_{m',M-1}\bs \epsilon_- + M \delta_{l',L}\delta_{m',M}\hat{\textbf{z}} ].
\end{align}
 However, more work is required for the first integral [Eq.~(\ref{eq:afirstint})]. Noting that
 \begin{align}
 \bt u \times \bs \epsilon_{\pm} = &\mp \im \cos \theta \bs \epsilon_\pm \pm \frac{\im}{2}e^{\pm \im \phi} \sin \theta \hat{ \bt z}, \\
 \bt u \times \hat{ \bt z}=& \sin \theta (\sin \phi \hat{\bt x} -\cos \phi \hat{ \bt y}) \nonumber \\
 =&-\im \sin\theta(e^{\im \phi} \bs \epsilon_- - e^{-\im \phi} \bs \epsilon_+),
 \end{align}
 we have,
 \begin{align}
a_{L,M}^{(l',m')}=\frac{\textbf{p}}{\sqrt{L(L+1)}}\cdot (  \bs \epsilon_+I_+ + \bs \epsilon_- I_- + \hat{ \bt z} I_0),
\end{align}
where
\bse
\label{eq:Iintegrals}
\begin{align}
I_0 =&  \int  \frac{\im}{2}\Big[c_{L,M}^{(-)}Y_{L,M+1}^* e^{\im \phi} -c_{L,M}^{(+)}Y_{L,M-1}^*e^{-\im \phi}\Big]  \nonumber \\
&\quad \times \sin \theta Y_{l',m'} \ud\Omega ,\\
I_\pm =& \int \Big[ \mp \im c_{L,M}^{(\mp)}Y_{L,M\pm1}^*\cos \theta Y_{l',m'} \nonumber \\
&\quad \pm \im M Y_{L,M}^*\sin\theta e^{\mp \im\phi} Y_{l',m'} \Big] \ud\Omega .
\end{align}
\ese
Using the following results:
 \begin{align}
 \int Y_{L,M}^*&e^{\im\phi}\sin \theta Y_{l',m'} d\Omega  \nonumber \\
 =&-\sqrt{\frac{(l'+m'+2)(l'+m'+1)}{(2l'+3)(2l'+1)}}\delta_{L,l'+1}\delta_{M,m'+1} \nonumber \\
 &+\sqrt{\frac{(l'-m')(l'-m'-1)}{4l'{}^2-1}}\delta_{L,l'-1}\delta_{M,m'+1},
 \end{align}
\begin{align}
  \int Y_{L,M}^*&e^{-\im \phi}\sin \theta Y_{l',m'} d\Omega  \nonumber \\
 =&-\sqrt{\frac{(l'+m')(l'+m'-1)}{4l'{}^2-1}}\delta_{L,l'-1}\delta_{M,m'-1} \nonumber \\
 &+\sqrt{\frac{(l'-m'+2)(l'-m'+1)}{(2l'+3)(2l'+1)}}\delta_{L,l'+1}\delta_{M,m'-1},
 \end{align}
\begin{align}
 \int Y_{L,M}^*&\cos \theta Y_{l',m'} d\Omega  \nonumber \\
 =&  \sqrt{\frac{(l'+1)^2-m'{}^2}{(2l'+3)(2l'+1)}}\delta_{L,l'+1}\delta_{M,m'} \nonumber \\&
 + \sqrt{\frac{l'{}^2-m'{}^2}{4l'{}^2-1}}\delta_{L,l'-1}\delta_{M,m'},
 \end{align}
we can solve the integrals in Eq.~(\ref{eq:Iintegrals}) and obtain
\begin{widetext}
\begin{align}
\label{eq:acoef}
a_{L,M}^{(l',m')}
=&\im \textbf{p} \cdot \Bigg\{ 
\sqrt{\frac{(L+1)}{L(2L+1)(2L-1)}} \Bigg[ 
-\sqrt{(L-M)(L-M-1)}\delta_{m',M+1}\bs \epsilon_+  + \sqrt{(L+M)(L+M-1)}\delta_{m',M-1}\bs \epsilon_-\nonumber \\
&
-\sqrt{(L-M)(L+M)}\delta_{m',M}\hat{ \bt z}
\Bigg]\delta_{l',L-1}
+\sqrt{\frac{L}{(L+1)(2L+1)(2L+3)}}  \Bigg[ 
-\sqrt{(L+M+2)(L+M+1)}\delta_{m',M+1}\bs \epsilon_+ \nonumber \\
&
+\sqrt{(L-M+2)(L-M+1)}\delta_{m',M-1}\bs \epsilon_-
+\sqrt{(L+M+1)(L-M+1)}\delta_{m',M}\hat{ \bt z}
\Bigg] \delta_{l',L+1}\Bigg\}.
\end{align}
\end{widetext}
Substituting Eq.~(\ref{eq:veinyz}) into Eq.~(\ref{eq:sveas}) gives,
\begin{align}
\textbf{V}_\textbf{r}^{(E)}\!\!\Lambda_{l,m}(\textbf{r}-\bs \rho_0) 
=&\sum_{L,M} \sum_{l',m'} \gamma_{l',m'}^{(l,m)}(\boldsymbol{\rho}_0) \\
&\times \left[ a_{L,M}^{(l',m')} \bs \Lambda_{L,M}^{(\text{I})}(\bt r) \right. 
 \left. + b_{L,M}^{(l',m')} \bs \Lambda_{L,M}^{(\text{II})}(\bt r) \right],  \nonumber
\end{align}
which in turn leads to the result in Eqs.~(\ref{eq:transtomult}) and (\ref{eq:coefftrans}) after using the results for the coefficients [Eqs.~(\ref{eq:bcoef}) and (\ref{eq:acoef})] and using the Kronecker deltas to eliminate one of the sums.


%

\end{document}